\documentclass{iopart}

\usepackage[german,english]{babel}
\usepackage{amssymb}
\usepackage{amsfonts}

\begin{document}

\letter{First passage and arrival time densities for L{\'e}vy flights
and the failure of the method of images}

\author{Aleksei V Chechkin\dag, Ralf Metzler\P, Vsevolod Y Gonchar\dag,
Joseph Klafter\ddag, and Leonid V Tanatarov\dag}
\address{\dag\ Institute for Theoretical Physics NSC KIPT,
Akademicheskaya st.1, 61108 Kharkov, Ukraine}
\address{\P\ NORDITA, Blegdamsvej 17, DK-2100 Copenhagen \O, Denmark}
\address{\ddag\ School of Chemistry, Tel Aviv University, 69978 Tel Aviv Israel}

\begin{abstract}
We discuss the first passage time problem in the semi-infinite interval, for
homogeneous stochastic Markov processes with L{\'e}vy stable jump length
distributions $\lambda(x)\sim\ell^{\alpha}/|x|^{1+\alpha}$ ($|x|\gg\ell$),
namely, L{\'e}vy flights (LFs). In particular, we demonstrate
that the method of images leads to a result, which violates a theorem due to
Sparre Andersen, according to which an arbitrary continuous and symmetric jump
length distribution produces a first passage time density (FPTD)
governed by the universal long-time decay $\sim t^{-3/2}$. Conversely, we
show that for LFs the direct definition known from Gaussian processes in
fact defines the probability density of first arrival, which for LFs differs
from the FPTD. Our findings are corroborated by numerical results.
\end{abstract}

\submitto{\JPA}

\vspace*{0.8cm}

L{\'e}vy flights (LFs) and L{\'e}vy walks (LWs) are the prime example in the
investigation of
non-standard transport processes whose stationary solution does not
converge towards the Boltzmann form \cite{bouchaud,report,frischbook,pt}.
Being subject to the generalised central limit theorem \cite{levy,gnedenko},
LFs correspond to a Markov process in which extremely long excursions can occur
with appreciable probability, whereas in LWs long excursions are penalised
through a time cost introduced via a spatiotemporal coupling
\cite{klablushle,some}.
Applications of LFs and LWs range from the famed
flight of an albatross \cite{stanley}, the spreading of spider-monkeys
\cite{spider}, or the grazing patterns of bacteria
\cite{levandowski}, over economical data \cite{mandelbrot} to molecular
collisions \cite{galgani} and plasmas \cite{chechkin1}. Despite their
broad usage, the detailed behaviour of even the simpler, uncoupled LF
processes, on which we concentrate in the following, in
external potentials and under non-trivial boundary conditions is still not
fully explored. Thus, there
have recently  been discovered bifurcations between multimodal states of
the probability density function (PDF) of LFs in steeper than harmonic
external fields, in whose presence also the variance becomes finite
\cite{chechkin,chechkina}, and rich band structures have
been reported for LFs in periodic potentials \cite{geisel}.

Of particular interest in random processes is the first passage time
density (FPTD) \cite{schroedinger,hughes,feller,redner}. For LFs, the
FPTD was determined through the method of images on a finite domain in
reference \cite{west}, and with similar methods in reference
\cite{gittermann}. These methods lead to results for the FPTD in the
semi-infinite domain, whose long-time behaviour explicitly depends on
the L{\'e}vy index $\alpha$. In contrast, a theorem due to Sparre Andersen
proves that for any discrete-time random walk process starting at $x_0
\neq 0$
with each step chosen from a continuous, symmetric but otherwise arbitrary
distribution, the FPTD asymptotically decays as $\sim n^{-3/2}$ with the
number $n$ of steps \cite{redner,sparre,sparre1}, being {\em fully
independent\/} of the index of the LF, i.e., universal. In the case of a
Markov process, the continuous time analogue of the Sparre Andersen result
reads
\begin{equation}
\label{universal}
p(t)\sim t^{-3/2}.
\end{equation}
The analogous universality was proved by Frisch and Frisch for the special
case in which an absorbing boundary is placed at the location of the
starting point of the LF at $t>0$ \cite{frisch}, and numerically
corroborated by Klafter and Zumofen \cite{zukla}. In the
following, we demonstrate that the method of images is generally
inconsistent with the universality of the FPTD, and therefore cannot be
applied to solve FPTD-problems for LFs. We also show that for LFs the FPTD
differs from the PDF for first arrival.

Let us start by recalling that an unbiased LF can be defined through
the space-fractional diffusion equation for the PDF $W(x,t)$
\cite{report,fogedby,sune}
\begin{equation}
\label{fde}
\frac{\partial}{\partial t}W=D\frac{\partial^{\alpha}}{\partial |x|^
{\alpha}}W(x,t)\therefore \quad \int_{-\infty}^{\infty}e^{\rmi kx}
\frac{\partial^{\alpha}W}{\partial |x|^{\alpha}}\rmd x\equiv -|k|^
{\alpha}W(k,t),
\end{equation}
where we define the fractional derivative $\partial^{\alpha}/\partial |x|^
{\alpha}$ by its Fourier transform. (Here and in the following, we restrict
ourselves to $1<\alpha<2$.) In position space, the fractional derivative
is defined in terms of the convolution (see \cite{chechkin} for the case
$\alpha=1$)
\begin{equation}
\label{riesz}
\frac{\partial^{\alpha}}{\partial |x|^{\alpha}}W(x,t)\equiv
\frac{D}{\kappa}
\frac{\partial^2}{\partial x^2}\int_{-\infty}^{\infty}
\frac{W(x',t)}{|x-x'|^{\alpha-1}}\therefore
\kappa\equiv 2\Gamma(2-\alpha)\left|\cos\frac{\pi\alpha}{2}\right|.
\end{equation}
Equivalently, LFs can be described in terms of
continuous time random walks with long-tailed jump length distributions
$\lambda(x)\sim\ell^{\alpha}/|x|^{1+\alpha}$ \cite{klablushle,hcf}. The
associated PDF $W(x,t)$ for natural boundary conditions ($\lim_{|x|\to
\infty}W(x,t)=0$) with initial condition $\delta(x)$ is the L{\'e}vy stable
law $W(x,t)=\int_{-\infty}^{\infty}\exp\left(-\rmi kx-D|k|^{\alpha}t\right)
\rmd k/(2\pi)$ \cite{levy,gnedenko}.
In Fourier-Laplace space \cite{REM}, this PDF corresponds to $W(k,s)=
\left(s+D|k|^{\alpha}\right)^{-1}$. A characteristic of LFs is the
divergence of the variance of both $W(x,t)$ and $\lambda(x)$. Equipping
equation (\ref{fde}) with a
$\delta$-sink of strength $p_{\rm fa}(t)$, we obtain the diffusion-reaction
equation for the non-normalised density function $f(x,t)$,
\begin{equation}
\label{heat}
\frac{\partial}{\partial t}f(x,t)=D\frac{\partial^{\alpha}}{\partial
|x|^{\alpha}}f(x,t)-p_{\rm fa}(t)\delta(x),
\end{equation}
from which by integration over the unrestricted space, we find the quantity
\begin{equation}
\label{surv}
p_{\rm fa}(t)=-\frac{\rmd}{\rmd t}\int_{-\infty}^{\infty}f(x,t)\rmd x,
\end{equation}
i.e., $p_{\rm fa}(t)$ is the negative time derivative of the survival
probability. By definition of the sink term, $p_{\rm fa}(t)$ is the PDF
of {\em first arrival\/}: once a random walker arrives at the sink, it
is annihilated. By solving equation (\ref{heat}) through standard methods
(determining the homogeneous and inhomogeneous solutions), it is
straightforward to calculate the solution $f$ in terms of the propagator
$W$ of equation (\ref{fde}) with initial condition $f(x,0)=\delta(x-x_0)$
yielding $f(k,u)=\left[e^{{\rm i}kx_0}+p_{\rm fa}(u)\right]/\left(s+D|k|^{
\alpha}\right)$, from which, in turn, we find that $p_{\rm fa}(t)$ satisfies
the chain rule ($p_{\rm fa}$ implicitly depending on $x_0$)
\begin{equation}
\label{mfpt}
W(-x_0,t)=\int_0^t p_{\rm fa}(\tau)W(0,t-\tau)\rmd \tau
\end{equation}
which corresponds to the Laplace space relation $p_{\rm fa}(s)=W(-x_0,s)/
W(0,s)$. Equation (\ref{mfpt}) is well-known and for any
sufficiently well-behaved continuum diffusion process is commonly employed
to define the FPTD \cite{hughes,redner}.

For Gaussian processes with propagator $W(x,t)=1/\sqrt{4\pi Dt}
\exp\left( -x^2/[4Dt]\right)$, one obtains by direct integration of the
diffusion equation with appropriate boundary condition the FPTD
\cite{redner}
\begin{equation}
\label{bmfpt}
p(t)=x_0(4\pi Dt^3)^{-1/2}\exp\left(-x_0^2/(4Dt)\right),
\end{equation}
including the asymptotic behaviour $p(t)\sim t^{-3/2}$ for $t\gg x_0^2/
(4D)$. In this Gaussian case, the quantity $p_{\rm fa}(t)$ is
equivalent to the FPTD. From a random walk perspective, this is due to
the fact that individual steps are of the same increment, and the jump length
statistics therefore ensures that the walker cannot hop across the sink
in a long jump without actually hitting the sink and being absorbed. This
behaviour becomes drastically different for L{\'e}vy jump length
statistics: There, the particle can easily cross the sink in a
long jump. Thus, before eventually being absorbed, it can pass by the
sink location numerous times, and therefore the statistics of the
first arrival will be different from the one of the first passage. In
fact, with $W(x,s)=(2\pi)^{-1}\int_{-\infty}^{\infty}e^{\rmi kx}\left(s+
D|k|^{\alpha}\right)^{-1}\rmd k$, we find
\begin{equation}
p_{\rm fa}(s)=1-\frac{\int_0^{\infty}\left(1-\cos kx_0\right)\big/\left(
s+Dk^{\alpha}\right)\rmd k}{\int_0^{\infty}1\big/\left(s+Dk^{\alpha}
\right)\rmd k}
\end{equation}
by use of the de Moivre identity $\exp(\rmi z)=\cos z+\rmi \sin z$. With
$\int_0^{\infty}(s+Dk^{\alpha})^{-1}\rmd k=\pi s^{1/\alpha-1}/(\alpha
D^{1/\alpha}\sin(\pi/\alpha))$ and
\[
\int_0^{\infty}\frac{1-\cos kx_0}{s+Dk^{\alpha}}\sim\frac{\Gamma((2-\alpha)
\sin\left(\pi (2-\alpha)/2\right)x_0^{\alpha-1}}{(\alpha-1)D},\quad s\to 0,
\alpha>1,
\]
we obtain the limiting form
\begin{equation}
\label{limit3}
p_{\rm fa}(s)\sim 1-x_0^{\alpha-1}s^{1-1/\alpha}D^{-1+1/\alpha}\tilde{
\Lambda}(\alpha),
\end{equation}
where $\tilde{\Lambda}(\alpha)=\alpha\Gamma(2-\alpha)\sin\left(\pi(2-
\alpha)/2\right)\sin(\pi/\alpha)/(\alpha-1)$. We note that the same
result is obtained by the exact expressions for $W(x_0,s)$ and $W(0,s)$
in terms of Fox $H$-functions and systematic expansion \cite{mathai}.
The inverse Laplace transform of the small $s$-behaviour (\ref{limit3})
can be obtained by completing (\ref{limit3}) to an exponential,
and then computing the Laplace inversion
by the identification $e^z=H^{1,0}_{0,1}[z|(0,1)]$ with the Fox
$H$-function \cite{mathai}, for which the exact Laplace inversion can be
performed \cite{wg}. Finally, a series expansion of this result leads to
the long-$t$ form
\begin{equation}
\label{limit4}
p_{\rm fa}(t)\sim C(\alpha)\frac{x_0^{\alpha-1}}{D^{1-1/\alpha}t^{2-1/
\alpha}},
\end{equation}
with $C(\alpha)=\alpha\Gamma(2-\alpha)\Gamma(2-1/\alpha)\sin\left(\pi[
2-\alpha]/2\right)\sin^2(\pi/\alpha)/(\pi^2(\alpha-1))$. Clearly, in the
Gaussian limit, the required asymptotic form $p(t)\sim x_0/\sqrt{4\pi D
t^3}$ for the FPTD is consistently recovered, whereas in the general case
the result (\ref{limit4}) is slower than in the universal FPTD behaviour
in equation (\ref{universal}), as it should as the $\delta$-trap used in
equation (\ref{heat}) to define the first arrival for LFs is weaker than the
absorbing wall used to properly define the FPTD. For LFs, the PDF for first
arrival thus scales like (\ref{limit4}) (i.e., it explicitly depends on the
index $\alpha$ of the underlying L{\'e}vy process), and, as shown below, it
{\em differs\/} from the corresponding FPTD.

\begin{figure}
\unitlength=1cm
\begin{picture}(10,7.2)
\put(1.8,8){\includegraphics{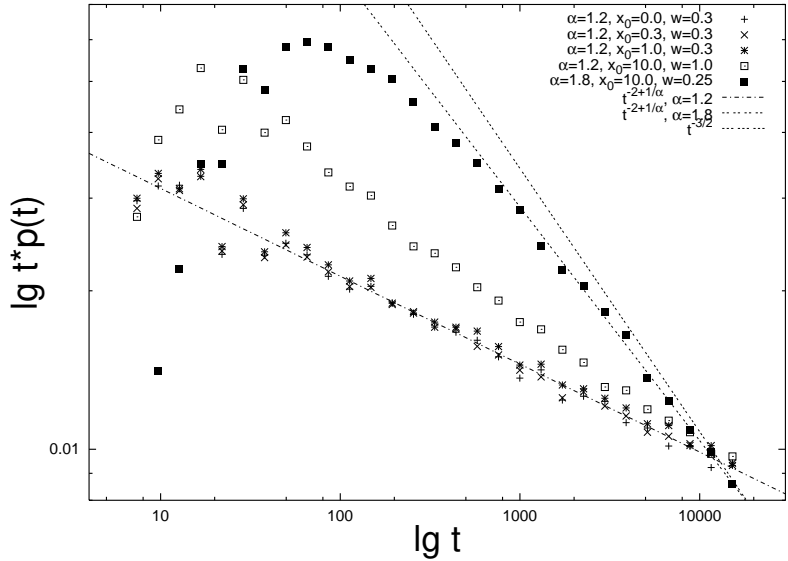}}
\end{picture}
\caption{First arrival PDF for $\alpha=1.2$ demonstrating the $t^{-2+1/
\alpha}$ scaling, for optimal trap width $w=0.3$. For comparison, we show
the same scaling for $\alpha=1.8$, and the power-law $t^{-3/2}$ corresponding
to the FPTD. The behaviour for too large $w=1.0$ shows a shift of the decay
towards the $-3/2$ slope. Note that on the abscissa we plot $\lg tp(t)$.
Note also that for the initial condition $x_0=0.0$, the trap becomes
activated {\em after\/} the first step, consistent with \protect\cite{zukla}.
\label{fig2}}
\end{figure}
Before calculating this FPTD, we first demonstrate the validity of equation
(\ref{limit4}) by means of a simulation the results of which are shown in
figure \ref{fig2}. Random jumps with LF jump length statistics are performed,
and a particle is removed when it hits a certain interval of width $w$ around
the sink; for our simulations we found an optimum value $w\approx 0.3$. As
seen in figure \ref{fig2} (note that we plot $\lg tp(t)$!) and for analogous
results not shown here, relation
(\ref{limit4}) is nicely fulfilled for all $1<\alpha<2$, whereas for larger
$w$, the slope increases.

\begin{figure}
\unitlength=1cm
\begin{picture}(10,7.2)
\put(0.94,-1.8){\includegraphics{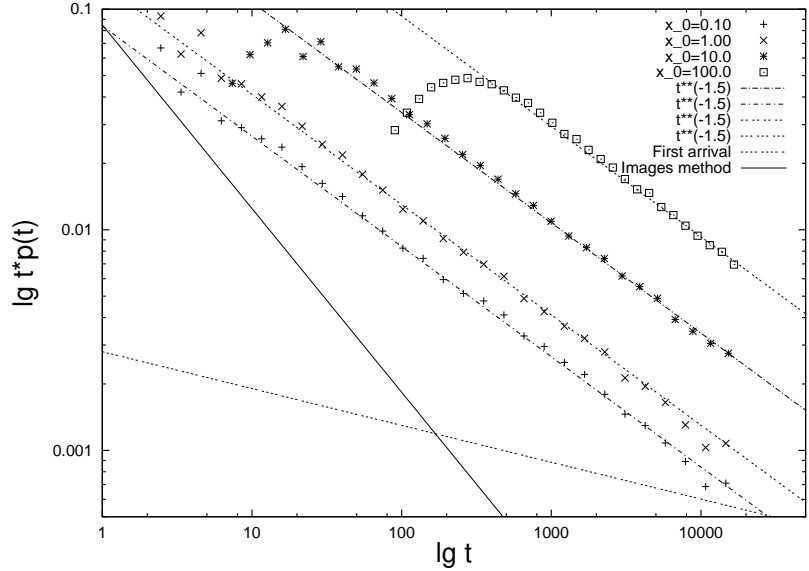}}
\end{picture}
\caption{Numerical results for the FPTD process on the semi-infinite
domain, for an LF with L{\'e}vy index $\alpha=1.2$. Note that on the
abscissa, we plot $tp(t)$. For all initial conditions $x_0=0.10$ 1.00,
10.0, and 100.0 the universal slope $-3/2$ in the $\log_{10}$-$\log_{10}$
plot is nicely reproduced, and it is significantly apart from the two
slopes predicted by the images method and the direct definition of the
FPTD.
\label{fig}}
\end{figure}

\begin{figure}
\unitlength=1cm
\begin{picture}(10,7.2)
\put(0.94,-1.8){\includegraphics{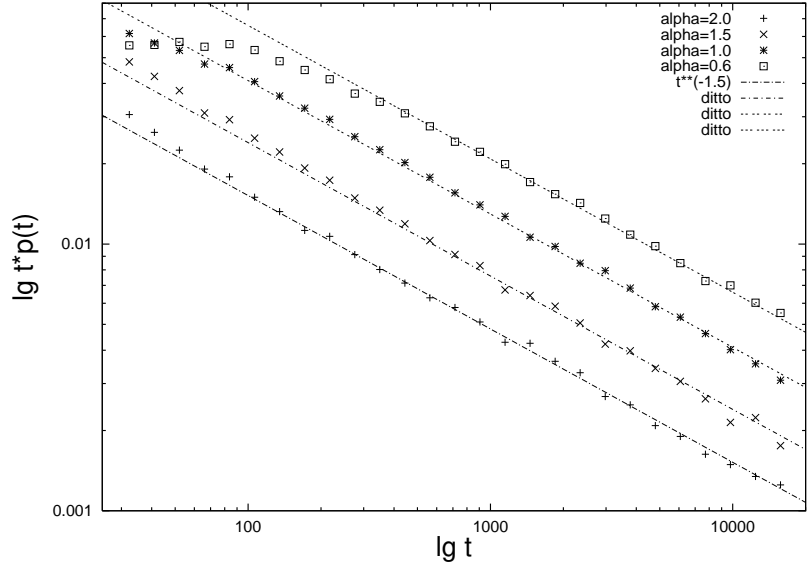}}
\end{picture}
\caption{Same as in figure \protect\ref{fig}, for $\alpha=2.0$, 1.5, 1.0,
and 0.6, and for the initial condition $x_0=10.0$. Again, the universal
$\sim t^{-3/2}$ behaviour is obtained.
\label{fig1}}
\end{figure}

The proper dynamical formulation of an LF on the semi-infinite interval
with an absorbing boundary condition at $x=0$, and thereby the
determination of the FPTD, has to make sure that in terms of above random
walk picture jumps across the sink are forbidden. This can be consistently
achieved by setting $f(x,t)\equiv 0$ on the left semi-axis, i.e.,
actually removing the particle when it crosses the point $x=0$.
This formally corresponds to the modified dynamical equation
\begin{equation}
\label{neu}
\frac{\partial f(x,t)}{\partial t}=\frac{D}{\kappa}\frac{\partial^2}
{\partial x^2}\int_0^{\infty}\frac{f(x',t)}{|x-x'|^{\alpha-1}}\rmd x'
\equiv\frac{\partial^2}{\partial x^2}{\cal F}(x,t)
\end{equation}
in which the fractional integral is truncated to the semi-infinite
interval. After Laplace transformation and integrating over $x$ twice,
one obtains
\begin{equation}
\int_0^{\infty}K(x-x',s)f(x',s)dx'=(x-x_0)\Theta(x-x_0)-xp(s)-{\cal F}(0,s),
\end{equation}
where $p(t)$ is the FPTD and the kernel $K(x,s)=sx\Theta(x)-(\kappa|x|^{
\alpha-1})$. This equation is formally of the Wiener-Hopf type of the
first kind \cite{gakhov}. After some manipulations similar to those
applied in reference \cite{zukla}, we arrive at the asymptotic expression
$p(s)\simeq 1-Cs^{1/2}$, where $C={\rm const}$,
in accordance with the universal behaviour (\ref{universal}) and with the
findings in reference \cite{zukla}. Thus, the
dynamic equation (\ref{neu}) consistently phrases the FPTD problem for
LFs. We note that due to the truncation of the fractional integral it was
not possible to modify the well-established Gr{\"u}nwald-Letnikov scheme
\cite{mainardi} to numerically solve equation (\ref{neu}) with enough
computational efficiency to numercially obtain the direct solution for
$f(x,t)$.
However, to corroborate the validity of the Sparre Anderson-universality,
we perform a simulation of an LF in the presence of an absorbing wall, i.e.,
random jumps with LF jump length statistics are performed along the right
semi-axis, and a particle is removed when it jumps across the origin to the
left semi-axis.
Results of such a detailed random walk study are displayed in figures \ref{fig}
and \ref{fig1}. The expected universal $t^{-3/2}$ scaling is
nicely obtained for various initial conditions and $\alpha$. Clearly, the
scaling for the first arrival as well as the image method-FPTD derived below
are significantly different.

We now demonstrate that the method of images produces a result, which is
neither consistent with the universal behaviour of the FPTD (\ref{universal})
nor with the behaviour of the PDF of first arrival (\ref{limit4}),
Given the initial condition $\delta(x-x_0)$, the solution $f_{\rm im}
(x,t)$ for the absorbing boundary value problem with the analogous
Dirichlet condition $f_{\rm im}(0,t)=0$ according to the method of images
is given via the difference \cite{feller,redner}
\begin{equation}
\label{im}
f_{\rm im}(x,t)=W(x-x_0,t)-W(x+x_0,t),
\end{equation}
in terms of the free propagator $W$, i.e., a {\em negative\/} image solution
originating at $-x_0$ balances the probability flux across the absorbing
boundary. The corresponding pseudo-FPTD is then calculated in the same way as
in equation (\ref{surv}). For the image solution in Fourier-Laplace
space, we obtain
\begin{equation}
f_{\rm im}(k,s)=[2\rmi \sin\left(kx_0\right)]/\left(s+D|k|^{\alpha}
\right),
\end{equation}
for a process which starts at $x_0>0$ and takes place in the right half
space. In Laplace space, the image method-FPTD becomes
\begin{equation}
p_{\rm im}(s)=1-s\int_0^{\infty}\rmd x \int_{-\infty}^{\infty}\frac{\rmd
k}{2\pi}e^{-\rmi kx}\frac{2\rmi \sin kx_0}{s+D|k|^{\alpha}}.
\end{equation}
After some transformations, we arrive at
\begin{equation}
\textstyle
p_{\rm im}(s)=1-2/\pi\int_0^{\infty}\rmd\xi\sin\left(\xi s^{1/\alpha}
x_0/D^{1/\alpha}\right)/\left[\xi\left(1+\xi^{\alpha}\right)\right].
\end{equation}
In the limit of small $s$, this expression reduces to
$p_{\rm im}(s)\sim 1-\Lambda(\alpha)x_0D^{-1/\alpha}s^{1/\alpha}$,
with $\Lambda(\alpha)=(2/\pi)\int_0^{\infty}\left(1+\xi^{\alpha}\right)
^{-1}\rmd\xi=2/(\alpha\sin(\pi/\alpha))$. Following the same procedure
as outlined above, we find the long-$t$ form
\begin{equation}
\label{limit2}
p_{\rm im}(t)\sim 2\Gamma(1/\alpha)x_0\Big/\left(\pi\alpha D^{1/\alpha}
t^{1+1/\alpha}\right)
\end{equation}
for the image method--FPTD. In the Gaussian limit $\alpha=2$, expression
(\ref{limit2}) produces $p_{\rm im}(t)\sim x_0/\sqrt{4\pi Dt^3}$, in
accordance to equation (\ref{bmfpt}). Conversely, for general $1<\alpha<2$,
$p_{\rm im}(t)$ according to equation (\ref{limit2}) would decay faster than $\sim
t^{-3/2}$. The method of images breaks down for LFs due to their special
non-local nature, displayed by the integrals in equations (\ref{fde}) and
(\ref{riesz}), and (\ref{neu}), namely having a long-tailed jump length
distribution. This leads to leapovers beyond the absorbing boundary.
The method of images is expected to work when the boundary is also a
turning poing of the trajectory, as actually happens for nearest
neighbours random walks, or the Wiener process.

Qualitatively, the following argument may be brought forth in favour of
the observed universality of the LF-FPTD: the long-time decay is expected
to be governed by short-distance jump events, corresponding to the central
region of very small jump lengths for the L{\'e}vy stable jump length
distribution.
But this region is, apart from a prefactor, indistinguishable from the
Gaussian distribution, and therefore the long-time behaviour should in
fact be the same for any continuous jump length distribution $\lambda(x)$.
In fact, the universal law (\ref{universal}) can only be modified in the
presence of non-Markov effects such as broad waiting time processes or
spatiotemporally coupled walks \cite{report,redner,klablushle,brian,trans}.
In terms of the special case covered by the theorem of Frisch and Frisch
\cite{frisch}, in which the absorbing boundary coincides with the initial
condition, we can understand the general situation for finite $x_0>0$, as
in the long-time limit, the distance $x_0$ becomes negligible in
comparison to the diffusion length $\langle|x (t)|\rangle\sim t^{1/
\alpha}$: therefore the asymptotic behaviour is necessarily
governed by the same universality.

Concluding, we demonstrated that the method of images, which has been
developed a powerful tool in Gaussian diffusion also beyond the 
homogeneous case \cite{feller,redner} and in the presence of long-tailed
waiting times \cite{report,brian,trans}, fails for
LF processes, leading to a false result for the FPTD. Moreover,
we showed that for such broad jump length statistics, the PDF of first
arrival at a point differs from the FPTD. We also provided a framework
in terms of a truncated fractional diffusion equation to solve the
FPTD problem for an LF. This study is expected to significantly contribute
to the understanding of the, at instances, non-trivial behaviour of LFs.

\ack

We are happy to acknowledge discussions with Igor Sokolov.
We also acknowledge financial support through the INTAS project 00-0847
from the European Commission.


\begin{thebibliography}{99}
\bibitem{REM} We denote the Laplace and Fourier transforms through the
explicit use of the image variables:
$p(s)\equiv\int_0^{\infty}p(t)e^{-st}\rmd t$, and $f(k,t)\equiv\int_{
-\infty}^{\infty}f(x,t)e^{\rmi kx}\rmd x.$
\end{thebibliography}
\end{document}